\documentclass[aps,prb,twocolumn,superscriptaddress]{revtex4-2}

\usepackage[breaklinks=true]{hyperref}
\hypersetup{colorlinks=true, citecolor=blue, urlcolor=blue, linkcolor=blue}

\usepackage[utf8]{inputenc}
\usepackage{graphicx}% Include figure files
\usepackage{dcolumn}% Align table columns on decimal point
\usepackage{bm}% bold math
\usepackage{amsmath}
\usepackage{amssymb}
\usepackage{cases}
\usepackage{diagbox}

\usepackage{color}

\let\p\partial
\let\f\frac
\let\t\textrm
\def\be{\begin{equation}}\def\ee{\end{equation}}
\def\bea{\begin{eqnarray}}\def\eea{\end{eqnarray}}

\newcommand{\B}[1]{\bm{#1}}

\begin{document}

\title{Boiling peak heat flux for steady inhomogeneous heat transfer in superfluid $^4$He}
\author{Sosuke Inui}
\affiliation{National High Magnetic Field Laboratory, 1800 East Paul Dirac Drive, Tallahassee, Florida 32310, USA}
\affiliation{Mechanical Engineering Department, FAMU-FSU College of Engineering, Florida State University, Tallahassee, Florida 32310, USA}

\author{Mikai Hulse}
\affiliation{National High Magnetic Field Laboratory, 1800 East Paul Dirac Drive, Tallahassee, Florida 32310, USA}
\affiliation{Department of Physics, Florida State University, Tallahassee, Florida 32310, USA}

\author{Toshiaki Kanai}
\affiliation{National High Magnetic Field Laboratory, 1800 East Paul Dirac Drive, Tallahassee, Florida 32310, USA}
\affiliation{Department of Physics, Florida State University, Tallahassee, Florida 32310, USA}

\author{Wei Guo}
\email[Corresponding: ]{wguo@magnet.fsu.edu}
\affiliation{National High Magnetic Field Laboratory, 1800 East Paul Dirac Drive, Tallahassee, Florida 32310, USA}
\affiliation{Mechanical Engineering Department, FAMU-FSU College of Engineering, Florida State University, Tallahassee, Florida 32310, USA}

\date{\today}
\begin{abstract}
Superfluid helium-4 (He II) is a widely adopted coolant in scientific and engineering applications owing to its exceptional heat transfer capabilities. However, boiling can spontaneously occur on a heating surface in He II when the heat flux exceeds a threshold value $q^*$, referred to as the peak heat flux. While the parameter $q^*$ holds paramount importance in the design of He II based cooling systems, extensive research has primarily focused on its behavior in steady homogeneous heat transfer from a flat heating surface. For inhomogeneous heat transfer from curved surfaces, $q^*$ exhibits intricate dependance on parameters such as the He II bath temperature $T_b$, the immersion depth $h$, and the curvature radius $R_0$ of the heating surface. A comprehensive understanding on how $q^*$ depends on these parameters remains elusive. In this paper, we report our systematic study on $q^*$ for steady heat transfer from cylindrical and spherical heaters in He II. We compute $q^*$ for a wide range of parameter combinations $(T_b, h, R_0)$ by solving the He II two-fluid equations of motion. The generated data have allowed us to develop a robust correlation that accurately reproduces $q^*$ for all the parameter combinations we explored. Our findings, particularly the establishment of the correlation, carry valuable implications for emergent applications that involve steady inhomogeneous heat transfer in He II systems.

\end{abstract}
\pacs{xxxx}
\maketitle

%%%%%%%%%%%%%%%%%%%%%%%%%%%%%%%%%%%%%%%%%%%%%%%%%%%%%%%%%%%%%%%%%%%%%%%%%%%%%
%%%%%%%%  Section 1  %%%%%%%%%%%%%%%%%%%%%%%%%%%%%%%%%%%%%%%%%%%%%%%%%%%%%%%%
%%%%%%%%%%%%%%%%%%%%%%%%%%%%%%%%%%%%%%%%%%%%%%%%%%%%%%%%%%%%%%%%%%%%%%%%%%%%%
\section{Introduction}
Saturated liquid $^4$He becomes a superfluid at temperatures below about 2.17 K~\cite{Tilley-book}. In the superfluid phase (known as He II), the liquid can be considered phenomenologically as a mixture of two miscible fluid components: an inviscid superfluid that carries no entropy and a viscous normal fluid that consists of thermal quasiparticles (i.e., phonons and rotons)~\cite{landau-1987}. Heat transfer in this two-fluid system is via a unique internal convection process known as thermal counterflow. In a counterflow, the normal fluid carries the heat and moves away from a heating surface at a velocity $v_n$=$q/\rho sT$, where $q$ is the heat flux, $T$ is the He II temperature, and $\rho$ and $s$ are the He II density and specific entropy, respectively; the superfluid moves in the opposite direction at a velocity $v_s$=$-v_n\rho_n/\rho_s$ so that the net mass flow remains zero (here $\rho_n$ and $\rho_s$ are the densities of the normal fluid and the superfluid, respectively). This counterflow mode is extremely effective, which renders He II a valuable coolant in a wide array of scientific and engineering applications, such as for cooling superconducting particle accelerator cavities, superconducting magnets, medical instruments, and even satellites \cite{vansciver-2012}.

When the relative velocity of the two fluids in counterflow exceeds a small critical value~\cite{Vinen-1957-PRS-I}, a chaotic tangle of quantized vortex lines can develop spontaneously in the superfluid. These quantized vortices are filamentary topological defects, each carrying a quantized circulation $\kappa\simeq10^{-3}$ cm$^2$/s around its angstrom-sized core~\cite{donnelly-1991}. A mutual friction force between the two fluids then emerges due to thermal quasiparticles scattering off the quantized vortices~\cite{vinen-1957_Proc.R.Soc.Lond.A_II}. This mutual friction can lead to novel flow characteristics in both fluids~\cite{marakov-2015_Phys.Rev.B, Gao-2017-PRB, Gao-2018-PRB, Bao-2018-PRB,Mastracci-2018-PRF}. When the heat flux is further increased to above a threshold value $q^*$, referred to as the peak heat flux, boiling on the heating surface can occur. This boiling action leads to the formation of vapor bubbles, and these bubbles can act as effective insulators between the heating surface and the surrounding He II, which impairs the heat transfer and results in the potential for overheating and damage to the cooled devices.

Developing a reliable correlation for assessing $q^*$ is of great importance in the design of He II based cooling systems. The value of $q^*$ can depend on many parameters, such as the heating duration $\Delta t$, the temperature of the He II bath $T_b$, the immersion depth $h$, and the curvature radius $R_0$ of the heating surface. In this paper, we shall focus on $q^*$ in steady heat transfer where $\Delta t\rightarrow\infty$, since this knowledge lays the groundwork for future explorations of $q^*$ within transient heat transfer scenarios.

There have been extensive studies on $q^*$ in the context of steady, homogeneous heat transfer of He II within uniform channels driven by planar heaters~\cite{fiszdon-1989_J.LowTemp.Phys.,fiszdon-1990_J.FluidMech.,shimazaki-1995_Cryogenics,hilton-2005_J.LowTemp.Phys.,zhang-2006_Int.J.HeatMassTransf.}. The relationship between $q^*$ and the parameters $T_b$ and $h$ has been reasonably well-understood~\cite{vansciver-2012}. However, when it comes to inhomogeneous heat transfer from curved surfaces such as cylindrical and spherical surfaces, $q^*$ displays intricate dependencies on the parameter combination $(T_b, h, R_0)$. Despite some past studies on $q^*$ for these nonuniform geometries~\cite{Lemieux67,Frederking68,Shiotsu1994,Goodling69,VanSciver80,Kryukov:2006tr}, a systematic understanding on how $q^*$ varies with the parameter combination $(T_b, h, R_0)$ remains absent. Nevertheless, establishing the capability to reliably predict $q^*$ values in these nonuniform geometries holds significant importance for specific applications, such as cooling superconducting transmission lines and magnet coils~\cite{maksoud-2010_IEEETrans.Appl.Supercond,xavier-2019_IEEETrans.Appl.Supercond}, detecting point-like quench spots on superconducting accelerator cavities~\cite{bao-2019_Phys.Rev.Applied, bao-2020_Int.J.HeatMassTransf.}, and emerging applications like the development of hot-wire anemometry for studying quantum turbulence in He II~\cite{duri-2015_Rev.Sci.Instrum}.

In this paper, we present a comprehensive numerical investigation of $q^*$ in steady, nonhomogeneous heat transfer from both cylindrical and spherical heating surfaces submerged in He II. We employ the He II two-fluid equations of motion to compute $q^*$ over a wide range of parameter combinations $(T_b, h, R_0)$. Furthermore, we demonstrate that the data we generate can facilitate the development of a robust correlation capable of accurately reproducing $q^*$ across all the parameter combinations we explore. The paper is structured as follows: we begin by outlining our theoretical model in Section~\ref{sec: Numerical Model}. In Section \ref{sec: Model Validation}, we conduct a comparative analysis of the calculated $q^*$ values for heat transfer from cylindrical heaters against available experimental data to calibrate our model. In Section~\ref{subsec: Cylindrical heater}, we present a systematic computation of $q^*$ using the fine-tuned model for cylindrical heaters under varying parameter combinations $(T_b, h, R_0)$ and establish a reliable correlation linking $q^*$ with these parameters. In Section~\ref{subsec: Spherical heater}, we provide a similar analysis and correlation for $q^*$ concerning heat transfer from spherical heaters. We conclude with a summary in Section~\ref{sec: Summary}.

%

%%%%%%%%%%%%%%%%%%%%%%%%%%%%%%%%%%%%%%%%%%%%%%%%%%%%%%%%%%%%%%%%%%%%%%%%%%%%%
%%%%%%%%  Section 2  %%%%%%%%%%%%%%%%%%%%%%%%%%%%%%%%%%%%%%%%%%%%%%%%%%%%%%%%
%%%%%%%%%%%%%%%%%%%%%%%%%%%%%%%%%%%%%%%%%%%%%%%%%%%%%%%%%%%%%%%%%%%%%%%%%%%%%
\section{Theoretical Model} \label{sec: Numerical Model}
We employ the two-fluid hydrodynamic model in our current research, which was also utilized in our prior work to analyze transient heat transfer in He II~\cite{SBaoTHTHeII,Sanavandi-PhysRevB.106.054501}. A comprehensive description of this model is available in Refs.~\cite{Nemirovskii-1995-RMP,nemirovskii2020closure}. In brief, this model is based on the conservation laws governing He II mass, momentum, and entropy. It comprises four evolution equations for He II's total density $\rho$, total momentum density $\rho \B{v}=\rho_s \B{v_s} + \rho_n \B{v_n}$, superfluid velocity $\B{v_s}$, and entropy $s$, as follows:
\begin{align}
  & {\f{\p \rho}{\p t}} + \nabla\cdot (\rho \B{v}) = 0, \label{eq: evol dinsity}\\
  & \f{\p (\rho\B{v})}{\p t} + \nabla (\rho_s v_s^2 + \rho_n v_n^2) + \nabla P= 0 , \label{eq: evol momentum}\\
  & \f{\p \B{v}_s}{\p t} + \B{v}_s\cdot \nabla \B{v}_s + \nabla \mu = \f{\B{F}_{ns}}{\rho_s} , \label{eq: evol super} \\
  & \f{\p (\rho s)}{\p t} + \nabla\cdot (\rho s  \B{v}_n) = \f{\B{F}_{ns}\cdot \B{v}_{ns}}{T}, \label{eq: evol entropy}
\end{align}
where $P$ is the pressure, $\mu$ is the chemical potential of He II, $\B{v}_{ns} = \B{v}_n-\B{v}_s$ is the relative velocity between two fluids, and $\B{F}_{ns}$ is the Gorter--Mellink mutual friction between the two fluids per unit volume of He II~\cite{vansciver-2012}.

$\B{F}_{ns}$ can be expressed in terms of $\B{v}_{ns}$ and the vortex-line density $L$ as \cite{Hall56a,Hall56b}:
\begin{equation} \label{eq: mutual friction}
	\B{F}_{ns} = \f{\kappa}{3}\f{\rho_s \rho_n}{\rho} B_L L \B{v}_{ns},
\end{equation}
where $B_L$ is a temperature dependent mutual friction coefficient~\cite{Donnelly-1998-JPCRD}. The calculation of $\B{F}_{ns}$ requires the evolution of $L(\B{r},t)$, for which we employ Vinen's equation~\cite{vinen-1957_Proc.R.Soc.Lond.A_II}:
\begin{equation} \label{eq: Vinen Eq}
	\f{\p L}{\p t} + \nabla \cdot ( \B{v}_{\t{\tiny L}} L) = \alpha_{\t{\tiny V}} |\B{v}_{ns}|L^{\f{3}{2}} - \beta_{\t{\tiny V}} L^2 + \gamma_{\t{\tiny V}}|\B{v}_{ns}|^{\f{5}{2}},
\end{equation}
where $\alpha_V$, $\beta_V$ and $\gamma_V$ are temperature-dependent empirical coefficients~\cite{vinen-1957_Proc.R.Soc.Lond.A_II}, and $\B{v}_{\t{\tiny L}}$ represents the vortex-tangle drift velocity, which is often approximated as equal to the local superfluid velocity $\B{v}_s$~\cite{schwarz-1988_Phys.Rev.B,nemirovskii-2019_LowTemp.Phys.}.

Similar to the previous works~\cite{SBaoTHTHeII,Sanavandi-PhysRevB.106.054501}, we include correction terms that depend on $v_{ns}^2$ for He II's thermodynamic properties, as suggested by Landau \cite{landau-1987,Khalatnikov-book}, to account for the large $v_{ns}$ values under high heat flux conditions in the current research:
\begin{align}
	&\mu(P,T,v_{ns}) = \mu^{(s)}(P,T) - \f{1}{2}\f{\rho_n}{\rho}v_{ns}^2, \label{eq: mu Landau}\\
	&s(P,T,v_{ns}) = s^{(s)}(P,T) + \f{1}{2}v_{ns}^2 \f{\p(\rho_n/\rho)}{\p T}, \\
	&\rho(P,T,v_{ns}) = \rho^{(s)}(P,T) + \f{1}{2} \rho^2 v_{ns}^2 \f{\p(\rho_n/\rho)}{\p P},
\end{align}
where the quantities with the superscript ``$^{(s)}$'' represent static values, which can be obtained from the HEPAK dynamic library \cite{hepak-2005}. The two-fluid model outlined above provides a coarse-grained description of the He II hydrodynamics, since it does not resolve the interaction between individual vortices and the normal fluid~\cite{Yui-2020-PRL, Mastracci-2019-PRF, Tang-NC-2023}. Nonetheless, prior research has shown that this model describes non-isothermal flows in He II well when $L$ is reasonably high~\cite{Sergeev-2019-EPL, SBaoTHTHeII}.

%%%%%%%%%%%%%%%%%%%%%%%%%%%%%%%%%%%%%%%%%%%%%%%%%%%%%%%%%%%
%%%%%%%%   Figure 1  %%%%%%%%%%%%%%%%%%%%%%%%%%%%%%%%%%%%%%
\begin{figure*}[t]
	\includegraphics [width=2\columnwidth]{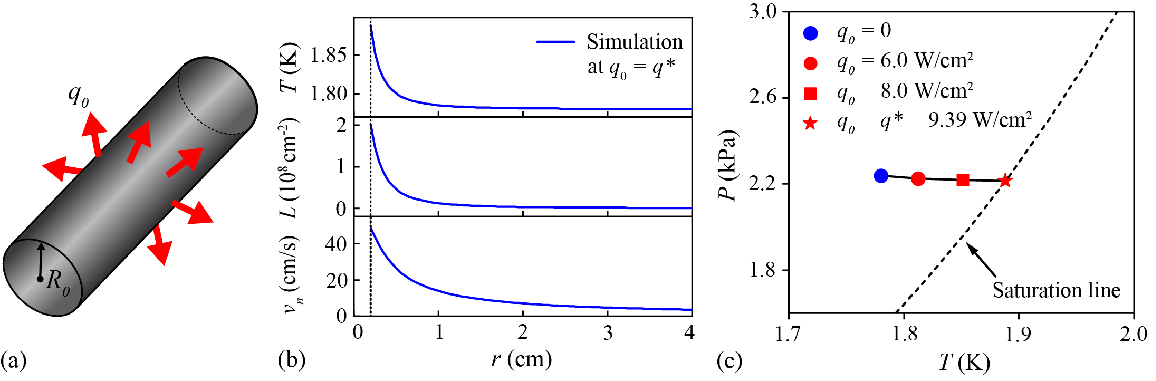}
	\caption{(a) A schematic diagram of a long cylindrical heater of radius $R_0$ with a constant surface heat flux $q_0$. (b) Simulated profiles of temperature $T(r)$ (top), vortex-line density $L(r)$ (middle), and normal-fluid velocity $v_n(r)$ (bottom) at $q_0$ close to 9.39~W/cm$^2$ with $T_b=1.78$~K, $R_0=0.2$~cm, and $h=50$~cm. (c) The calculated state parameter $(P_0, T_0)$ of the He II on the heater surface at various applied $q_0$.
}
	\label{fig: 1}
\end{figure*}
%%%%%%%%%%%%%%%%%%%%%%%%%%%%%%%%%%%%%%%%%%%%%%%%%%%%%%%%%%

Since our current research focuses on the steady-state heat transfer, we drop the terms that involve the time derivative in the governing equations and reformulate them in a manner convenient for numerical solutions. For instance, Eq.~\eqref{eq: evol dinsity} leads to:
\be
\rho_s \B{v}_s = -\rho_n \B{v}_n \label{eq: identity from density}.
\ee
Moreover, by integrating Eq.~\eqref{eq: evol momentum}, we can derive an expression for $P(r)$ as:
\be \label{eq: P integral}
P(r) = P_b -  \rho_s v_s^2 - \rho_n v_n^2.
\ee
Here the bath pressure $P_b=P_S(T_b) + \rho g h$, where $P_S(T_b)$ represents the saturation pressure at the bath temperature $T_b$, $g$ stands for gravitational acceleration, and $h$ denotes the immersion depth of the heating surface. The last two terms in Eq.~(\ref{eq: P integral}) account for the Bernoulli pressures associated with the flows in the two fluids. Now, assuming axial symmetry and recognizing the identity $d\mu=\frac{1}{\rho}dP-sdT$~\cite{landau-1987}, we can express Eq.~\eqref{eq: evol super} in the following form, utilizing Eqs.~\eqref{eq: evol momentum} and \eqref{eq: identity from density}:
\begin{equation} \label{eq: steady super}
  \rho v_{s} \f{\p v_n}{\p r} + \p_r (\rho {v}_{n}{v}_{s} ) =   \rho s \f{\p T}{\p r} + \f{\rho}{\rho_s}{F}_{ns}.
\end{equation}
The temperature $T(r)$ at location $r$ can be obtained by integrating the above equation as:
\be \label{eq: T integral}
T(r) = T_{b} + \int_r^{\infty} \t{d}r' G(r'),
\ee
where
\be
G(r) := \f{1}{\rho_s s}{F}_{ns}- \f{{v}_{s}}{s} \p_r {v}_n - \f{1}{\rho s} \p_r (\rho {v}_{n}{v}_{s} ).
\ee
Next, we integrate Eq.~\eqref{eq: evol entropy} from the heater surface $R_0$ to $r$ to obtain $v_n (r)$ as:
\begin{equation} \label{eq: v_n integral}
	v_n(r) = \f{R_0^N \rho_0 s_0}{r^N \rho s}v_{n0} + \f{1}{r^N \rho s}I(r),
\end{equation}
where
\begin{equation}
	I(r) := \int_{R_0}^{r} \t{d}r' r'^N \f{F_{ns}(r')v_{ns}(r')}{T(r')}.
\end{equation}
The quantities with the subscript ``$_0$'' in the above equations indicate their values at $r=R_0$, and the parameter $N$ assumes values of 1 or 2, corresponding to cylindrical and spherical coordinates, respectively. Note that $v_{n0}$ is related to the surface heat flux $q_0$ as $v_{n0} = q_0/\rho_0 s_0 T_0$, which transforms Eq.~\eqref{eq: v_n integral} into:
\be  \label{eq: v_n integral modified}
v_n(r) = \f{ q_0}{\rho s T_0}\left(\f{R_0}{r}\right)^N  + \f{I(r)}{r^N \rho s}.
\ee
Finally, within the parameter ranges explored in our current research, it becomes evident that the drift term $\nabla \cdot ( \B{v}_{\t{\tiny L}} L)$ and the term $\gamma_{\t{\tiny V}}|\B{v}_{ns}|^{\f{5}{2}}$ in Eq.~(\ref{eq: Vinen Eq}) are orders of magnitudes smaller than the remaining terms. By omitting these two terms, we can deduce that $L(r) = \gamma^2 v_{ns}(r)^2$, where $\gamma=\alpha_V / \beta_V$. Therefore, $F_{ns}$ can be calculated as:
\begin{equation}
F_{ns}= \f{\kappa}{3}\f{\rho_s \rho_n}{\rho} B_L \gamma^2 v_{ns}^3. \label{eq: F_ns-final}
\end{equation}
We must emphasize that Eq.~(\ref{eq: Vinen Eq}) was originally proposed for homogeneous and isotropic counterflow. There are ongoing discussions regarding potential modifications of this equation for nonuniform flows~\cite{MONGIOVI20181,Khomenko:2015uy,Nemirovskii:2018tf}. In our present research, we will maintain the use of Eq.~(\ref{eq: F_ns-final}). However, we will adapt the $\gamma$ values, originally derived for uniform counterflow~\cite{Childers,Tough82,Adachi-2010-PRB}, to best fit the available data under nonuniform counterflow conditions. The relevant details are provided in Sec.~\ref{sec: Model Validation}.

Eqs.~(\ref{eq: identity from density}), (\ref{eq: P integral}), (\ref{eq: T integral}), (\ref{eq: v_n integral modified}), and (\ref{eq: F_ns-final}) now form the base of our iterative numerical approach for solving the steady-state heat transfer problems involving cylindrical and spherical heaters. The iteration starts with constant He II properties $P^{(0)}=P_b$ , $T^{(0)}=T_b$, and a prescribed normal-fluid velocity profile $v_n^{(0)}(r)=\f{ q_0}{\rho^{(0)} s^{(0)} T_0}\left(\f{R_0}{r}\right)^N$. Here, the superscript $^{(i=0,1,2...)}$  denotes the iteration number. Utilizing the initial fields $(P^{(0)},T^{(0)}, v_n^{(0)})$, we can calculate all relevant He II thermodynamic variables and other needed parameters, such as $v_s^{(0)}$, $\rho^{(0)}$, $s^{(0)}$, $F_{ns}^{(0)}$, etc. These results allow us to iteratively update $(P,T,v_n)$ as:
\begin{align} \label{eq: all update}
P^{(i+1)}(r) &=P_b +  \rho_s^{(i)} v_s^{(i)}(r)^2 + \rho_n^{(i)} v_n^{(i)}(r)^2, \\
T^{(i+1)}(r) &= T_b + \int_{\infty}^r \t{d}r' G^{(i)}(r'), \\	
v_n^{(i+1)}(r) &= \f{ q_0}{\rho^{(0)} s^{(0)} T_0}\left(\f{R_0}{r}\right)^N  + \f{I^{(i)}(r)}{r^N \rho^{(i)} s^{(i)}}.
\end{align}
The iteration is terminated once the relative change in the temperature field between consecutive iterations, defined as $|T^{(i)}-T^{(i-1)} |/T^{(i)}$, becomes less than $10^{-5}$ at all $r$. In the simulation, the integrals are performed using Simpson's rule with a step size of $\Delta r=10$~$\mu$m~\cite{riley_hobson_bence_2006}.

As an example, we consider a cylindrical heater with a radius $R_0 = 0.2$~cm, subject to a constant surface heat flux $q_0$, as depicted in Fig.~\ref{fig: 1}(a). We set $T_b = 1.78$~K and $h = 50$~cm, and compute the steady-state profiles of $T(r)$, $L(r)$, and $v_n(r)$ using the iterative method outlined earlier. The results for $q_0$ close to 9.39~W/cm$^2$ are shown in Fig.~\ref{fig: 1}(b). It is clear that approaching the heater, $T(r)$, $L(r)$, and $v_n(r)$ all increase rapidly towards their maximum values at $r=R_0$. In Fig.~\ref{fig: 1}(c), we show the state parameters $(T_0,P_0)$ of the He II on the heater surface at various $q_0$. The blue dot represents the state $(T_0=T_b,P_0=P_b)$ at $q_0=0$. As $q_0$ increases, the state approaches the saturation line of He II. The slight reduction in pressure is due to the Bernoulli effect incorporated in Eq.~(\ref{eq: P integral}). At the peak heat flux $q^*\approx9.39$~W/cm$^2$, the He II state on the heater surface reaches the saturation line, where boiling can occur spontaneously.

%%%%%%%%%%%%%%%%%%%%%%%%%%%%%%%%%%%%%%%%%%%%%%%%%%%%%%%%%%%%%%%%%%%%%%%%%%%%%
%%%%%%%%  Section 3  %%%%%%%%%%%%%%%%%%%%%%%%%%%%%%%%%%%%%%%%%%%%%%%%%%%%%%%%
%%%%%%%%%%%%%%%%%%%%%%%%%%%%%%%%%%%%%%%%%%%%%%%%%%%%%%%%%%%%%%%%%%%%%%%%%%%%%
\section{Model calibration} \label{sec: Model Validation}
To calibrate our model, we have looked into existing experimental research on $q^*$ associated with steady-state nonuniform heat transfer in He II. There were several experimental studies on $q^*$ for cylindrical heaters~\cite{Lemieux67,Frederking68,Shiotsu1994,Goodling69,Irey:1975vp,VanSciver80}. As for spherical heaters, research has been limited, primarily focusing on transient heat transfer scenarios or heat flux magnitudes considerably lower than $q^*$~\cite{Kryukov:2006tr,Xie:2022up}. Among the available studies on $q^*$ for cylindrical heaters, several studies employed thin-wire heaters with radii in the range of $1 \sim 10^2$ $\mu$m~\cite{Lemieux67,Frederking68,Shiotsu1994}, which is comparable to the mean vortex-line spacing $\ell=L^{-1/2}$ observed in such experiments. We choose to avoid those particular datasets in our study, since our coarse-gained model is applicable only at length scales much greater than $\ell$. In subsequent analyses, we will focus on comparing our numerical simulation results with the data reported in Refs.~\cite{vansciver-2012,VanSciver80, VanSciver81}, where cylindrical heaters of notable diameters were used.

%%%%%%%%%%%%%%%%%%%%%%%%%%%%%%%%%%%%%%%%%%%%%%%%%%%%%%%%%%
%%%%%%%%  Figure 2  %%%%%%%%%%%%%%%%%%%%%%%%%%%%%%%%%%%%%%
\begin{figure}[t!]
	\includegraphics [width=1.0\columnwidth]{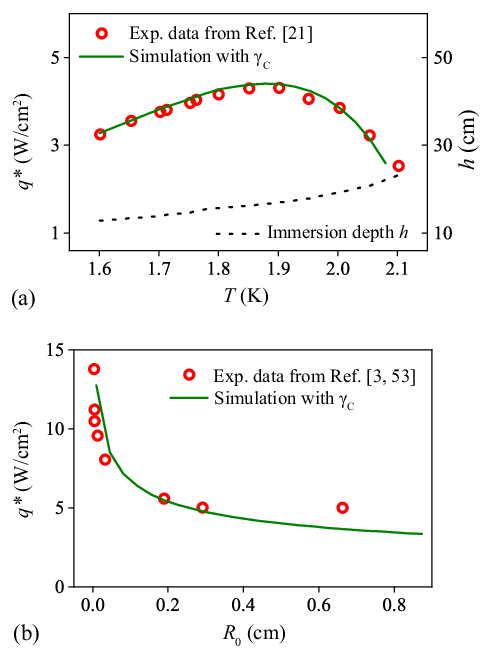}
	\caption{(a) Comparison of simulated peak heat flux $q^*$ at different He II bath temperature $T_b$ with experimental data from Ref.~\cite{VanSciver80} for a cylindrical heater with $R_0=0.66$~cm. The heater immersion depth $h$ (dashed curve) varied with $T_b$ in the experiment. (b) Simulated dependence of $q^*$ on heater radius $R_0$ at $h=10$~cm and $T_b=1.78$~K, compared with experimental data from Ref.~\cite{vansciver-2012,VanSciver81}. $\gamma_C=1.2\gamma$ was adopted in these simulations.
}
    \label{fig: 2}
\end{figure}
%%%%%%%%%%%%%%%%%%%%%%%%%%%%%%%%%%%%%%%%%%%%%%%%%%%%%%%%%%

\begin{figure}[t!]
	\includegraphics [width=0.9\columnwidth]{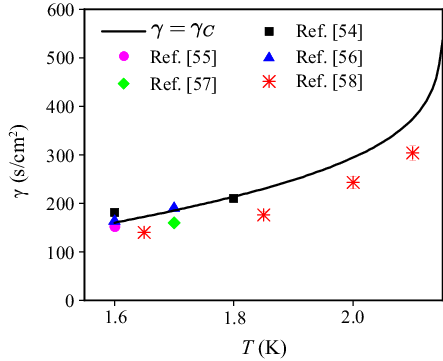}
	\caption{The optimized $\gamma_C$ as a function of $T$. Measured $\gamma$ values in uniform counterflow in various experiments \cite{Chase-PhysRev.127.361,Dimotakis-phys_fluids,Martin-PhysRevB.27.2788,Babuin-PhysRevB.86.134515,Gao-PhysRevB.96.094511} are also shown.  }
    \label{fig: 3}
\end{figure}

Fig.~\ref{fig: 2}(a) presents the measured $q^*$ values at different He II bath temperatures $T_b$ for a cylindrical heater with $R_0=0.66$~cm~\cite{VanSciver80}. In the referenced experiment, the immersion depth $h$ declined as the bath was pumped to achieve lower $T_b$. Consequently, distinct $T_b$ values correspond to varying $h$ levels, as illustrated by the dotted curve in Fig.~\ref{fig: 2}(a). Our model simulations have taken this variation into account. Fig.~\ref{fig: 2}(b) displays the measured $q^*$ values for heaters with different radii $R_0$ at $T_b=1.78$~K and $h=10$~cm \cite{vansciver-2012,VanSciver81}. To compare with these data, we have calculated $q^*$ under identical conditions using our iterative method. In our calculations, we adopted the $\gamma$ coefficient derived from the He II heat conductivity function $f(T)$ under saturated vapor pressure, defined as~\cite{vansciver-2012}:
\begin{equation}
	f(T) = \f{A_{GM}\rho_n}{\rho_s^3 s^4 T^3},
\end{equation}
where $A_{GM} \approx \f{1}{3} \f{\rho_n}{\rho}B_L \kappa \gamma^2$ is the Gorter--Melink mutual friction coefficient \cite{donnelly-1991}. The expressions of $f$ and $A_{GM}$ lead to the following identity for $\gamma$:
\begin{equation}
	\gamma = \sqrt{ \f{3 \rho_s^3 \rho s^4 T^3 f}{\rho_n \kappa B_L}  }. \label{eq: gamma}
\end{equation}
Experimentally, the temperature-dependence of $f(T)$ in uniform counterflow has been studied thoroughly, and its values are compiled in Ref.~\cite{vansciver-2012}. Therefore, the value of $\gamma$ for uniform counterflow can be easily calculated. However, when it comes to nonuniform counterflow, there is little knowledge on how $\gamma$ may change. In this context, we opt to scale the $\gamma$ values deduced from Eq.~(\ref{eq: gamma}) by a factor $C$, yielding $\gamma_C=C\gamma$. We treat $C$ as an adjustable parameter. Remarkably, with $C=1.2$, our simulation results (illustrated as solids curves in Fig.~\ref{fig: 2}) exhibit excellent agreement with the experimental data. The optimized $\gamma_C$ as a function of $T$ is shown in Fig.~\ref{fig: 3} together with some $\gamma$ values obtained in uniform counterflow experiments. In the subsequent sections, we will apply the optimized $\gamma_C$ in our systematic analysis of $q^*$.

%%%%%%%%%%%%%%%%%%%%%%%%%%%%%%%%%%%%%%%%%%%%%%%%%%%%%%%%%%
%%%%%%%%  Figure 3  %%%%%%%%%%%%%%%%%%%%%%%%%%%%%%%%%%%%%%
%%%%%%%%%%%%%%%%%%%%%%%%%%%%%%%%%%%%%%%%%%%%%%%%%%%%%%%%%%

%%%%%%%%%%%%%%%%%%%%%%%%%%%%%%%%%%%%%%%%%%%%%%%%%%%%%%%%%%%%%%%%%%%%%%%%%%%%%
%%%%%%%%  Section 4  %%%%%%%%%%%%%%%%%%%%%%%%%%%%%%%%%%%%%%%%%%%%%%%%%%%%%%%%
%%%%%%%%%%%%%%%%%%%%%%%%%%%%%%%%%%%%%%%%%%%%%%%%%%%%%%%%%%%%%%%%%%%%%%%%%%%%%
\section{Peak Heat Flux Analysis}
In this section, we present the simulated $q^*$ values for steady-state counterflow produced by both cylindrical and spherical heaters, considering a variety of parameter combinations $(T_b, h, R_0)$. We further demonstrate that $q^*$ can be calculated using an integral formula that involves the temperature difference between the heater surface and the bath. Using our simulation data, we can devise a correlation to evaluate this temperature difference, which in turn leads to a robust correlation for $q^*$.

%%%%%%%%%%%%%%%%%%%%%%%%%%%%%%%%%%%%%%%%%%%%%%%%%%%%%%%%%%%%%%%%%%%%%%%%%%%%%
%%%%%%%%  Subsection A  %%%%%%%%%%%%%%%%%%%%%%%%%%%%%%%%%%%%%%%%%%%%%%%%%%%%%
%%%%%%%%%%%%%%%%%%%%%%%%%%%%%%%%%%%%%%%%%%%%%%%%%%%%%%%%%%%%%%%%%%%%%%%%%%%%%
\subsection{Cylindrical heater case} \label{subsec: Cylindrical heater}
\begin{figure}[t!]
	\includegraphics [width=0.9\columnwidth]{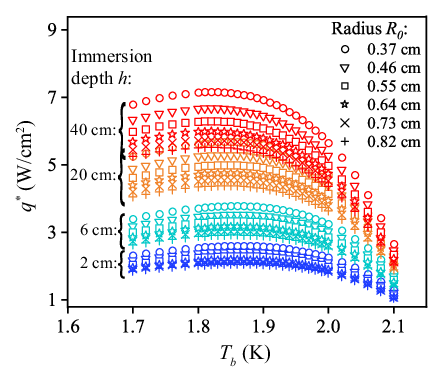}
	\caption{Simulated peak heat flux $q^*$ for cylindrical heaters with various $T_b$, $h$, and $R_0$.}
    \label{fig: 4}
\end{figure}

Following the same procedures as illustrated in Fig.~\ref{fig: 1}, we determined $q^*$ as a function of $T_b$ for cylindrical heaters of various $R_0$ and $h$ values. These results are compiled in Fig.~\ref{fig: 4}. It's evident that at fixed $R_0$ and $h$ values, $q^*$ exhibits a non-monotonic dependence on $T_b$, with a peak observed between 1.8~K and 1.9~K. On the other hand, at a fixed $T_b$, $q^*$ consistently increases with an increase in $h$ or a decrease in $R_0$.

To understand the behavior of $q^*$, we can refer to Eq.~\eqref{eq: steady super}. For the parameter combinations $(T_b, h, R_0)$ that we studied, we found that the terms on the left-hand side of Eq.~\eqref{eq: steady super} are typically more than two orders of magnitude smaller than the other terms across all values of $r$. If we dismiss these minor terms and utilize Eq.~\eqref{eq: F_ns-final} and \eqref{eq: gamma}, while noting that $v_{ns}(r) = q(r)/\rho_s s T$, the following equation can be derived:
\begin{equation}
	\f{\t{d} T}{\t{d} r}  =  -C^2 f(T) q(r)^3.
\end{equation}
In steady-state counterflow, $q(r)$ is given by $q(r)=q_0(R_0/r)^N$ (recall that $N=1$ for cylindrical heaters and $N=2$ for spherical heaters). When the heater surface heat flux $q_0$ reaches $q^*$, the above equation can be rearranged and integrated to produce an expression for $q^*$:
\begin{equation} \label{eq: peak heat flux integral}
	q^* = \left( \f{3N - 1}{C^2 R_0}  \int_{T_b}^{T_b+\Delta T} \f{\t{d} T}{f(T)} \right)^{1/3},
\end{equation}
where $\Delta T$ denotes the temperature increase on the heater surface relative to the He II bath at $q_0=q^*$. This equation was introduced in Ref.~\cite{vansciver-2012}. However, due to the lack of information on how $\Delta T$ depends on $(T_b, h, R_0)$, this equation was not employed to evaluate $q^*$.

%%%%%%%%%%%%%%%%%%%%%%%%%%%%%%%%%%%%%%%%%%%%%%%%%%%%%%%%%%
%%%%%%%%  Figure 4  %%%%%%%%%%%%%%%%%%%%%%%%%%%%%%%%%%%%%%
%%%%%%%%%%%%%%%%%%%%%%%%%%%%%%%%%%%%%%%%%%%%%%%%%%%%%%%%%%

\begin{figure}[t!]
	\includegraphics [width=0.9\columnwidth]{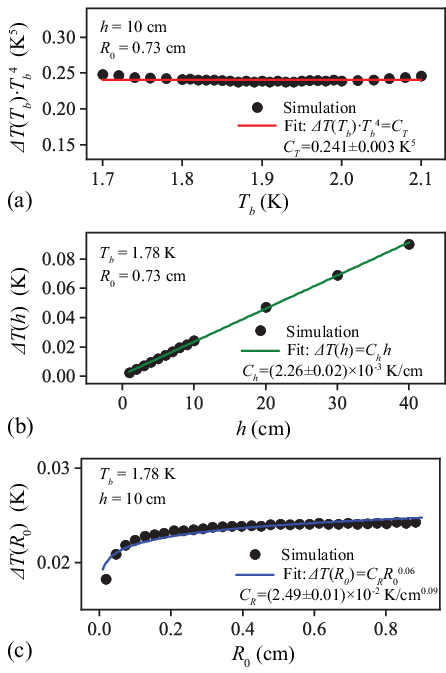}
	\caption{(a) Simulated temperature rise $\Delta T$ on the heater surface as a function of $T_b$ for a cylindrical heater with fixed $h$ and $R_0$. (b) Dependence of $\Delta T$ on the immersion depth $h$ at fixed $T_b$ and $R_0$. (c) Dependence of $\Delta T$ on the heater radius $R_0$ at fixed $T_b$ and $h$.}
    \label{fig: 5}
\end{figure}

To facilitate the development of a practical correlation for $q^*$, we have computed $\Delta T$ values for all the cases depicted in Fig.~\ref{fig: 5}. Some results showing relationship of $\Delta T$ with $T_b$, $h$, $R_0$ are presented in panels (a), (b), and (c) of Fig.~\ref{fig: 5}. From Fig.~\ref{fig: 5}(a), we can see that at fixed $h$ and $R_0$, $\Delta T$ largely scales as $T_b^{-4}$ across the entire bath temperature range we explored. Fig.~\ref{fig: 5}(b) demonstrates a rather good linear dependence of $\Delta T$ on $h$ for given $T_b$ and $R_0$. Lastly, Fig.~\ref{fig: 5}(c) reveals a somewhat mild power-law dependance, $\Delta T\propto R_0^{\alpha}$, when $T_b$ and $h$ are fixed. This power exponent $\alpha$ varies with $h$ and $T_b$, as listed in Table \ref{tab: alpha cylinder}, and is generally small. Combining all these insights, we can propose the following simple correlation between $\Delta T$ and the parameters $T_b$, $h$, and $R_0$:
\begin{equation}  \label{eq: delta T for cylinder}
	\Delta T(T_b,h,R_0,) = D \f{h R_0^{\alpha}}{T_b^4},
\end{equation}
where $D$ is a numerical factor derivable from the scaling coefficients shown in Fig.~\ref{fig: 5}(a)-(c). To evaluate $D$ in a more systematic manner, we compute it as $D=\Delta T/(hR_0^\alpha /T_b^4)$ for each parameter combination $(T_b, h, R_0)$. Notably, within our chosen parameter range, all deduced values for $D$ fall within the range $D = 0.024 \pm 0.002$ K$^5$/cm$^{1+\alpha}$. More details regarding the derivation of $D$ is provided in Appendix~\ref{Appendix A}.

%%%%%%%%%%%%%%%%%%%%%%%%%%%%%%%%%%%%%%%%%%%%%%%%%%%%%%%%%%
%%%%%%%%  Figure 5  %%%%%%%%%%%%%%%%%%%%%%%%%%%%%%%%%%%%%%
%%%%%%%%%%%%%%%%%%%%%%%%%%%%%%%%%%%%%%%%%%%%%%%%%%%%%%%%%%

%%%%%%%%%%%%%%%%%%%%%%%%%%%%%%%%%%%%%%%%%%%%%%%%%%%%%%%%%%
%%%%%%%%  Table I  %%%%%%%%%%%%%%%%%%%%%%%%%%%%%%%%%%%%%%%
\begin{table}[h!]
	\caption{\label{tab: alpha cylinder}The fitted exponent $\alpha$ for cylindrical heaters}
	\begin{tabular}{|c||*{4}{c|}}\hline
		\backslashbox{$h$ [cm]}{$T_b$ [K]}
		&\makebox[4em]{1.7}&\makebox[4em]{1.8}&\makebox[4em]{1.9}&\makebox[4em]{2.0}\\\hline\hline
		1  &0.12&0.11&0.09&0.07\\\hline
		5  &0.08&0.07&0.06&0.04\\\hline
		20 &0.05&0.04&0.03&0.02\\\hline
	\end{tabular}
\end{table}
%%%%%%%%%%%%%%%%%%%%%%%%%%%%%%%%%%%%%%%%%%%%%%%%%%%%%%%%%%

\begin{figure}[t!]
	\includegraphics [width=0.9\columnwidth]{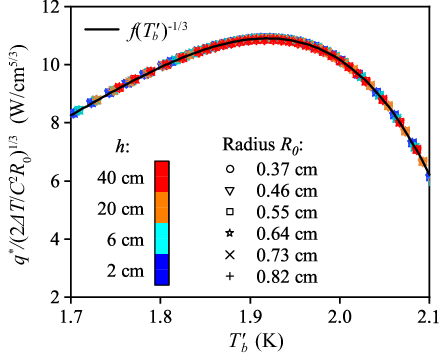}
	\caption{Simulated $q^*/(2\Delta T/C^2 R_0)^{1/3}$ as a function of $T_b'=T_b+\Delta T/2$ for cylindrical heaters at all the parameter combinations $(T_b, h, R_0)$ we studied. The black curve represents $f^{-1/3}(T_b')$, where $f$ is the known He II heat conductivity function~\cite{vansciver-2012}. The simulated data collapse nicely onto the $f^{-1/3}$ curve.}
    \label{fig: 6}
\end{figure}

With the obtained expression for $\Delta T$, we can now derive a convenient correlation to evaluate $q^*$. Given that $\Delta T$ is typically much smaller than $T_b$ (i.e., see Fig.~\ref{fig: 5}), the integral in Eq.~\eqref{eq: peak heat flux integral} can be approximated by evaluating $f(T)$ at $T = T_b +\f{1}{2}\Delta T$, resulting in:
\begin{equation}  \label{eq: peak heat flux approx}
	q^* \approx   \left((3N-1)\Delta T/C^2 R_0\right)^{1/3}\cdot f(T_b+ \Delta T/2)^{-1/3}.
\end{equation}
To verify the accuracy of this expression for cylindrical heaters, we plot the simulated $q^*/(2\Delta T/C^2 R_0)^{1/3}$ in Fig.~\ref{fig: 6} as a function of $T_b'=T_b+\Delta T/2$ for all the parameter combinations we studied. Impressively, all the simulated data collapse onto a single curve, which agrees precisely with $f(T_b')^{-1/3}$.

%%%%%%%%%%%%%%%%%%%%%%%%%%%%%%%%%%%%%%%%%%%%%%%%%%%%%%%%%%
%%%%%%%%  Figure 6  %%%%%%%%%%%%%%%%%%%%%%%%%%%%%%%%%%%%%%
%%%%%%%%%%%%%%%%%%%%%%%%%%%%%%%%%%%%%%%%%%%%%%%%%%%%%%%%%%

In order to derive a convenient correlation for $q^*$ that explicitly depends on $T_b$, $h$, and $R_0$, one can perform a Taylor expansion of Eq.~\eqref{eq: peak heat flux approx} as:
\begin{equation}  \label{eq: peak heat flux approx2}
	q^* \approx   (2\Delta T/C^2 R_0)^{1/3} \left[   \frac{1}{f(T_b)} -  \f{\Delta T}{2} \f{f'(T_b)}{f(T_b)^2}   \right]^{1/3}.
\end{equation}
Using the expression for $\Delta T$ from Eq.~\eqref{eq: delta T for cylinder}, we can substitute it into Eq.~\eqref{eq: peak heat flux approx2} to yield the following final correlation:
\begin{equation}  \label{eq: peak heat flux formula cylinder}
q^* \approx   \left[ \f{2 D  h }{C^2 R_0^{1-\alpha}T_b^4 f(T_b)} \left(   1 -  \f{D  h R_0^{\alpha}}{2T_b^4} \f{f'(T_b)}{f(T_b)}   \right) \right]^\f{1}{3}.	
\end{equation}
With this correlation, evaluating $q^*$ becomes straightforward given a specific set of parameters $(T_b, h, R_0)$. It is worth noting from Eq.~\eqref{eq: peak heat flux formula cylinder} that the dependance of $q^*$ on $R_0$ can be expressed as $q^*\propto R_0^{-\frac{1}{m}}$, where $m \approx 3 / (1-\alpha)$. For the parameter ranges explored in our simulations, $m$ varies from 3.06 to 3.4. The deviation of $m$ from 3 is entirely due to the weak dependance of $\Delta T$ on $R_0$, i.e., $\Delta T\propto R_0^\alpha$ as shown in Eq.~\eqref{eq: delta T for cylinder}. It is worth highlighting that such a deviation from $m=3$ has indeed been reported experimentally~\cite{vansciver-2012}.

%%%%%%%%%%%%%%%%%%%%%%%%%%%%%%%%%%%%%%%%%%%%%%%%%%%%%%%%%%%%%%%%%%%%%%%%%%%%%
%%%%%%%%  Subsection B  %%%%%%%%%%%%%%%%%%%%%%%%%%%%%%%%%%%%%%%%%%%%%%%%%%%%%
%%%%%%%%%%%%%%%%%%%%%%%%%%%%%%%%%%%%%%%%%%%%%%%%%%%%%%%%%%%%%%%%%%%%%%%%%%%%%
\subsection{Spherical heater case} \label{subsec: Spherical heater}
\begin{figure}[t!]
	\includegraphics [width=0.9\columnwidth]{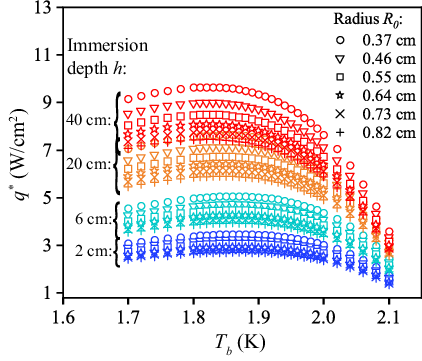}
	\caption{Simulated peak heat flux $q^*$ for spherical heaters with various $T_b$, $h$, and $R_0$.}
    \label{fig: 7}
\end{figure}

In the case of spherical heaters, we follow a similar procedure to that for the cylindrical heaters. We consider a spherical heater of radius $R_0$ immersed at depth $h$ in He II held at a bath temperature $T_b$, and then conduct numerical simulations across various $T_b$, $h$, and $R_0$ values. The obtained $q^*$ data are displayed in Fig.~\ref{fig: 7}. From the data, it's evident that the variation of $q^*$ with respect to $T_b$, $h$, and $R_0$ for spherical heaters show similar trends observed for cylindrical heaters. Moreover, for a given parameter set $(T_b, h, R_0)$, the $q^*$ value for spherical heaters is consistently higher than that for cylindrical heaters.
%%%%%%%%%%%%%%%%%%%%%%%%%%%%%%%%%%%%%%%%%%%%%%%%%%%%%%%%%%
%%%%%%%%  Figure 7  %%%%%%%%%%%%%%%%%%%%%%%%%%%%%%%%%%%%%%
%%%%%%%%%%%%%%%%%%%%%%%%%%%%%%%%%%%%%%%%%%%%%%%%%%%%%%%%%%

\begin{figure}[t!]
	\includegraphics [width=0.9\columnwidth]{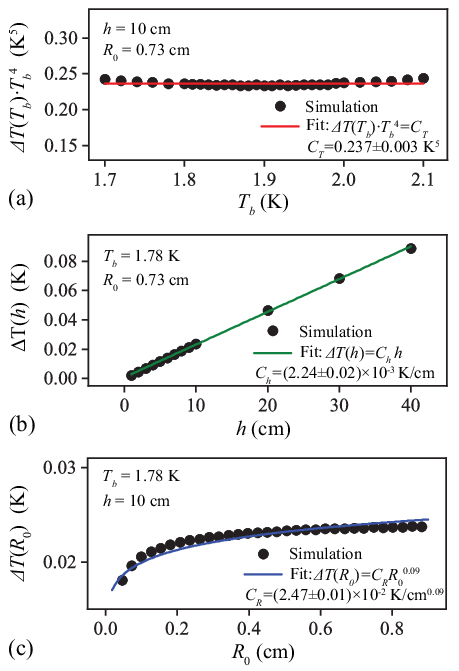}
	\caption{(a) Dependence of $\Delta T$ on the bath temperature $T_b$ for a spherical heater of fixed $h$ and $R_0$. (b) Dependence of $\Delta T$ on the immersion depth $h$ at fixed $T_b$ and $R_0$. (c) Dependence of $\Delta T$ on the heater radius $R_0$ at fixed $T_b$ and $h$. }
    \label{fig: 8}
\end{figure}

The behavior of $\Delta T$ for spherical heaters closely mirrors what we observed for cylindrical heaters. In Fig.~\ref*{fig: 8}(a), \ref*{fig: 8}(b), and \ref*{fig: 8}(c), we display representative results showing the dependencies of $\Delta T$ on $T_b$, $h$, and $R_0$. These results lead us to a correlation for $\Delta T$ which strikingly takes the same form as Eq.~\eqref{eq: delta T for cylinder} for cylindrical heaters, namely $\Delta T=D(h R_0^\alpha/T_b^4)$. The fitted values of $\alpha$ (as shown in Table~\ref{tab: alpha spherical}) is approximately double that of cylindrical heaters. The similarity of these expressions underscores the robustness of the correlation across different heater geometries. As before, the factor $D$ for each parameter set $(T_b,h,R_0)$ can be computed as $D=\Delta T/(hR_0^{\alpha} /T_b^4)$. The resulting values of $D$ for all studied cases fall within $D = 0.024 \pm 0.002$ K$^5$/cm$^{1+\alpha}$, matching precisely with those derived for cylindrical heaters. Further details on the derivation of $D$ is provided in Appendix~\ref{Appendix A}.

%%%%%%%%%%%%%%%%%%%%%%%%%%%%%%%%%%%%%%%%%%%%%%%%%%%%%%%%%%
%%%%%%%%  Figure 8  %%%%%%%%%%%%%%%%%%%%%%%%%%%%%%%%%%%%%%
%%%%%%%%%%%%%%%%%%%%%%%%%%%%%%%%%%%%%%%%%%%%%%%%%%%%%%%%%%

%%%%%%%%%%%%%%%%%%%%%%%%%%%%%%%%%%%%%%%%%%%%%%%%%%%%%%%%%%
%%%%%%%%  Table II  %%%%%%%%%%%%%%%%%%%%%%%%%%%%%%%%%%%%%%
\begin{table}[t!]
	\caption{\label{tab: alpha spherical}The fitted exponent $\alpha$ for spherical heaters}
	\begin{tabular}{|c||*{4}{c|}}\hline
		\backslashbox{$h$ [cm]}{$T_b$ [K]}
		&\makebox[4em]{1.7}&\makebox[4em]{1.8}&\makebox[4em]{1.9}&\makebox[4em]{2.0}\\\hline\hline
		1  &0.20&0.18&0.15&0.11\\\hline
		5  &0.13&0.11&0.10&0.07\\\hline
		20 &0.08&0.07&0.06&0.04\\\hline
	\end{tabular}
\end{table}
%%%%%%%%%%%%%%%%%%%%%%%%%%%%%%%%%%%%%%%%%%%%%%%%%%%%%%%%%%

\begin{figure}[t!]
	\includegraphics [width=0.90\columnwidth]{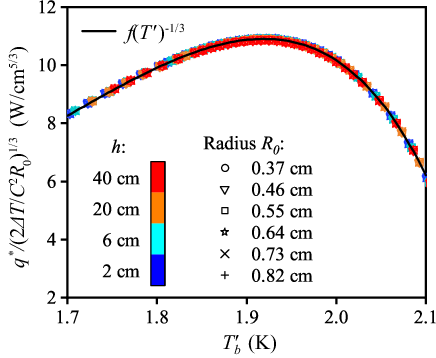}
	\caption{Simulated $q^*/(2\Delta T/C^2 R_0)^{1/3}$ as a function of $T_b'=T_b+\Delta T/2$ for spherical heaters at all the parameter combinations $(T_b, h, R_0)$ we studied. The black curve represents $f^{-1/3}(T_b')$, where $f$ is the known He II heat conductivity function~\cite{vansciver-2012}.}
    \label{fig: 9}
\end{figure}

To demonstrate the precision of Eq.~\eqref{eq: peak heat flux approx} for spherical heaters, we again plot $q^*/(5\Delta T/C^2 R_0)^{1/3}$ against $T^\prime_b = T_b + \f{1}{2}\Delta T$. As shown in Fig. \ref{fig: 9}, data points for all parameter combinations $(T_b, h, R_0)$ collapse onto a single curve descried by $f(T^\prime_b)^{-1/3}$. Finally, using a similar approach, we can express $q^*$ for spherical heaters explicitly in terms of $T_b$, $h$ and $R_0$ by incorporating the expression for $\Delta T$:
\begin{equation}  \label{eq: peak heat flux spherical}
	q^* \approx   \left[ \f{ 5 D h }{C^2 R_0^{1-\alpha}T_b^4 f(T_b)} \left(   1 -  \f{D h R_0^{\alpha}}{2T_b^4} \f{f'(T_b)}{f(T_b)}   \right) \right]^\f{1}{3}.
\end{equation}
Compared to Eq.~\eqref{eq: peak heat flux formula cylinder}, apart from the variance in $\alpha$, the main difference lies in the numerical factor $3N-1=5$ for the spherical geometry.

%%%%%%%%%%%%%%%%%%%%%%%%%%%%%%%%%%%%%%%%%%%%%%%%%%%%%%%%%%
%%%%%%%%  Figure 9  %%%%%%%%%%%%%%%%%%%%%%%%%%%%%%%%%%%%%%
%%%%%%%%%%%%%%%%%%%%%%%%%%%%%%%%%%%%%%%%%%%%%%%%%%%%%%%%%%

\section{Summary}  \label{sec: Summary}
We have conducted a comprehensive numerical analysis of the boiling peak heat flux $q^*$ for steady-state heat transfer in He II from both cylindrical and spherical heaters. The $q^*$ value was calculated using the He II two-fluid equations of motion for given bath temperature $T_b$, heater immersion depth $h$, and heater radius $R_0$. We calibrated our model by comparing the simulated $q^*$ values with available experimental data under the same parameter combinations $(T_b, h, R_0)$. The optimized model was then utilized to generate $q^*$ values across a wide parameter range. Based on the obtained data, we developed convenient correlations of $q^*$ that explicitly depend on $(T_b, h, R_0)$ for both cylindrical and spherical heaters. Notably, while spherical heaters generally exhibit higher $q^*$ values than their cylindrical counterparts under identical parameters, the derived correlations share a structural resemblance. These correlations are valuable in the design of cooling systems that involve steady but inhomogeneous heat transfer in He II. Looking ahead, we plan to extend the current work to evaluate $q^*$ in transient heat transfer of He II in nonhomogeneous geometries. For such transient heat transfer, the correlation of $q^*$ is expected to be more complicated, since it will depend not only on $(T_b, h, R_0)$ but also the heating duration $\Delta t$. The insights obtained in the current research will form the foundation for our future transient heat transfer analysis.

\begin{acknowledgments}
The authors acknowledge the support by the US Department of Energy under Grant DE-SC0020113 and the Gordon and Betty Moore Foundation through Grant GBMF11567. The work was conducted at the National High Magnetic Field Laboratory at Florida State University, which is supported by the National Science Foundation Cooperative Agreement No. DMR-2128556 and the state of Florida.
\end{acknowledgments}

\appendix
\section{Determination of $D$ factor}  \label{Appendix A}
\begin{figure}[b]
	\includegraphics [width=0.9\columnwidth]{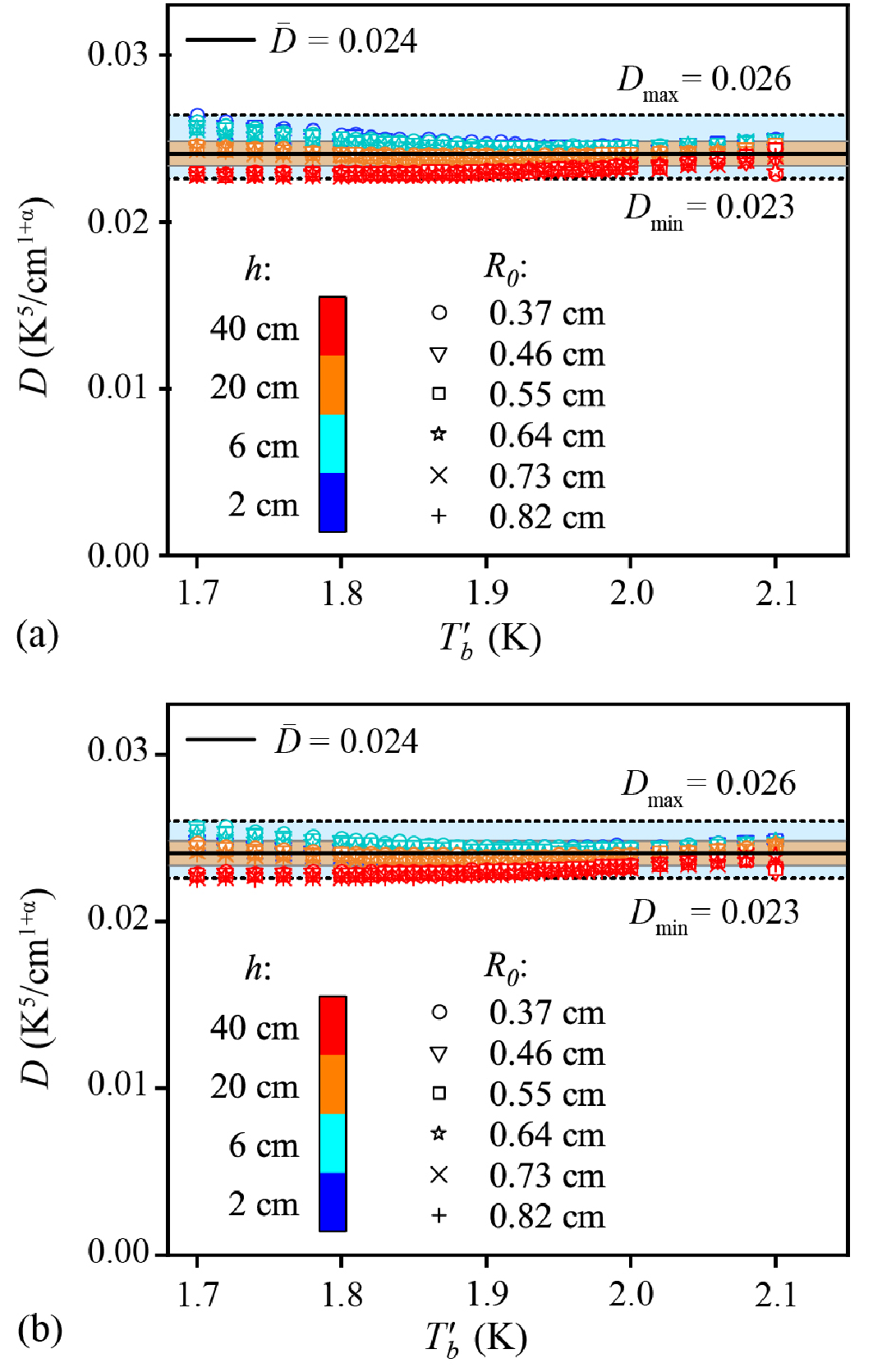}
	\caption{(a)--(b) Correlation factor $D$ for cylindrical and spherical heaters, respectively, calculated under all the parameter combinations $(T_b, h, R_0)$ we explored.}
	\label{fig: 10}
\end{figure}

In the main text, we discussed that the temperature rise $\Delta T=T_0-T_b$ at the peak heat flux $q^*$ can be expressed in terms of the bath temperature $T_b$, the hydrostatic head $h$, and the heater radius $R_0$ as given by Eq.~\eqref{eq: delta T for cylinder}. To determine $D$ in a systematic manner, we calculate it as $D=\Delta T/(hR_0^\alpha /T_b^4)$ for each parameter combination $(T_b, h, R_0)$. Fig.~\ref{fig: 10} (a) and (b) show the results for cylindrical and spherical heaters, respectively. The data cover a wide range of $T_b$, $h$ and $R_0$ and are indicated by distinct marker shapes and colors. It is clear that $D$ remains roughly constant across all the parameter combinations. In each figure, two colored bands are shown. The narrow band shown in orange represents the region bounded by $D = \bar{D} \pm \sigma_D$, where $\bar{D}=0.024$ K$^5$/cm$^{1+\alpha}$ is the mean value of $D$ averaged over all the data points and $\sigma_D$ denotes the standard deviation. The wide band shown in blue is bounded by the maximum $D_\t{max} = 0.026$ K$^5$/cm$^{1+\alpha}$ and the minimum $D_\t{min} = 0.023$ K$^5$/cm$^{1+\alpha}$ among all the data points. It is clear that all the $D$ values fall within the range $D = 0.024 \pm 0.002$ K$^5$/cm$^{1+\alpha}$, across the parameter ranges considered in the paper, for both cylindrical and spherical heaters.

%%%%%%%%%%%%%%%%%%%%%%%%%%%%%%%%%%%%%%%%%%%%%%%%%%%%%%%%%%%
%%%%%%%%   Figure 10  %%%%%%%%%%%%%%%%%%%%%%%%%%%%%%%%%%%%%%
%%%%%%%%%%%%%%%%%%%%%%%%%%%%%%%%%%%%%%%%%%%%%%%%%%%%%%%%%%

%\bibliographystyle{junsrt}
%\bibliography{REFERENCE_books, REFERENCE_papers}
\bibliography{Ref}

\end{document}